# The X-ray counterpart to the gravitational wave event GW 170817

E. Troja[1,2], L. Piro[3], H. van Eerten[4], R. T. Wollaeger[5], M. Im[6], O. D. Fox[7], N. R. Butler[8], S. B. Cenko[2,9], T. Sakamoto[10], C. L. Fryer[5], R. Ricci[11], A. Lien[2,12], R. E. Ryan Jr.[7], O. Korobkin[5], S.-K. Lee[6], J. M. Burgess[13], W. H. Lee[14], A. M. Watson[14], C. Choi[6], S. Covino[15], P. D'Avanzo[15], C. J. Fontes[5], J. Becerra González[16,17], H. G. Khandrika[7], J. Kim[6], S. -L. Kim[18], C. -U. Lee[18], H. M. Lee[19], A. Kutyrev[1,2], G. Lim[6], R. Sánchez-Ramírez[3], S. Veilleux[1,9], M. H. Wieringa[20], Y. Yoon[6]

[1]Department of Astronomy, University of Maryland, College Park, MD 20742-4111, USA

[2]Astrophysics Science Division, NASA Goddard Space Flight Center, 8800 Greenbelt Rd, Greenbelt, MD 20771, USA

[3]INAF, Istituto di Astrofisica e Planetologia Spaziali, via Fosso del Cavaliere 100, 00133 Rome, Italy

[4]Department of Physics, University of Bath, Claverton Down, Bath BA2 7AY, United Kingdom

[5]Center for Theoretical Astrophysics, Los Alamos National Laboratory, Los Alamos, NM 87545 USA

[6]Center for the Exploration for the Origin of the Universe, Astronomy Program, Dept. of Physics & Astronomy, Seoul National University, 1 Gwanak-ro, Gwanak-gu, Seoul 08826, Republic of Korea

[7]Space Telescope Science Institute, Baltimore MD, 21218

[8]School of Earth and Space Exploration, Arizona State University, Tempe, AZ 85287, USA

[9]Joint Space-Science Institute, University of Maryland, College Park, MD 20742, USA

[10]Department of Physics and Mathematics, Aoyama Gakuin University, 5-10-1 Fuchinobe, Chuo-ku, Sagamihara-shi Kanagawa 252-5258, Japan

[11]INAF-Istituto di Radioastronomia, Via Gobetti 101, I-40129, Italy

[12]Department of Physics, University of Maryland, Baltimore County, 1000 Hilltop Circle, Baltimore, MD 21250, USA

[13]Max-Planck-Institut für extraterrestrische Physik, Giessenbachstrasse, D-85748 Garching, Germany

[14]Instituto de Astronomía, Universidad Nacional Autónoma de México, Apartado Postal 70-264, 04510 México, CDMX, Mexico

[15]INAF/Brera Astronomical Observatory, via Bianchi 46, Merate (LC), Italy

[16]Inst. de Astrofísica de Canarias, E-38200 La Laguna, Tenerife, Spain

[17]Universidad de La Laguna, Dpto. Astrofísica, E-38206 La Laguna, Tenerife, Spain

[18]Korea Astronomy and Space Science Institute, 776 Daedeokdae-ro, Yuseong-gu, Daejeon 34055, Korea

[19]Astronomy Program, Dept. of Physics & Astronomy, Seoul National University, 1 Gwanak-ro, Gwanak-gu, Seoul 08826, Republic of Korea

[20]CSIRO Astronomy and Space Science, P.O. Box 76, Epping NSW 1710, Australia

**A long-standing paradigm in astrophysics is that collisions- or mergers- of two neutron stars (NSs) form highly relativistic and collimated outflows (jets) powering gamma-ray bursts (GRBs) of short (< 2 s) duration[1,2,3]. However, the observational support for this model is only indirect[4,5]. A hitherto outstanding prediction is that gravitational wave (GW) events from such mergers should be associated with GRBs, and that a majority of these GRBs should be off-axis, that is, they should point away from the Earth[6,7]. Here we report the discovery of the X-ray counterpart associated with the GW event GW170817. While the electromagnetic counterpart at optical and infrared frequencies is dominated by the radioactive glow from freshly synthesized r-process material in the merger ejecta[8,9,10], known as kilonova, observations at X-ray and, later, radio frequencies exhibit the behavior of a short GRB viewed off-axis[7,11]. Our detection of X-ray emission at a location coincident with the kilonova transient provides the missing observational link between short GRBs and GWs from NS mergers, and gives independent confirmation of the collimated nature of the GRB emission.**

On 17 August 2017 at 12:41:04 Universal Time (UT; hereafter $T_0$), the Advanced Laser Interferometer Gravitational-Wave Observatory (LIGO) detected a gravitational wave transient from the merger of two NSs at a distance of 40 +/- 8 Mpc[12]. Approximately two seconds later, a weak gamma-ray burst (GRB) of short duration (<2 s) was observed by the *Fermi* Gamma-ray Space Telescope[13] and INTEGRAL[14]. The low-luminosity of this gamma-ray transient was rather unusual when compared to the population of short GRBs at cosmological distances[15], and its physical connection with the GW event remained unclear.

A vigorous observing campaign targeted the localization region of the GW transient, and rapidly identified a source of bright optical, infrared (IR), and ultraviolet (UV) emission in the early-type galaxy NGC 4993[16,17]. This source, designated SSS17a, was initially not visible at radio and X-ray wavelengths. However, on 26 Aug 2017, we observed the field with the *Chandra X-ray Observatory* and detected X-ray emission at the position of SSS17a (Figure 1). The observed X-ray flux (see Methods) implies an isotropic luminosity of 9 x $10^{38}$ erg $s^{-1}$ if located in NGC 4993 at a distance of ~40 Mpc. Further *Chandra* observations, performed between 01 and 02 Sep 2017, confirmed the presence of continued X-ray activity, and hinted at a slight increase in luminosity to $L_{X,iso}$ ~ 1.1 x $10^{39}$ erg $s^{-1}$. At a similar epoch the onset of radio emission was also detected[18].

The evolution of SSS17a across the electromagnetic spectrum shows multiple components dominating the observed emission. Simple modeling of the optical-infrared photometry as a black body in linear expansion suggests mildly relativistic ($\geq 0.2c$) velocities and cool (<10,000K) temperatures. We find a hot blue component, mainly contributing at optical wavelengths, and a colder infrared component, which progressively becomes redder (Extended Data Figure 1). The low peak luminosity ($M_V$~-16) and featureless optical spectrum (Figure 2) disfavour a supernova explosion (see Methods), while the broad ($\Delta\lambda/\lambda \approx 0.1$) features in the IR spectra are consistent with expectations for rapidly expanding dynamical ejecta[9,10], rich in lanthanides and actinides. The overall properties of the host galaxy, such as its stellar mass, evolved stellar population and low star formation (see Methods), are consistent with the typical environment of short GRBs and in line with the predictions for compact binary mergers[5]. When combined, these data point to a kilonova emission, consisting of the superposition of radioactive-powered emission from both neutron-rich dynamical ejecta expanding with velocity $v$ ~ $0.2c$ and a slower, sub-relativistic wind[19]. The former component radiates most of its energy in the IR, while the latter dominates the optical and UV spectrum. The optical/IR dataset therefore provides convincing evidence that SSS17a was a kilonova produced by the merger of two compact objects, at a time and location consistent with GW170817.

Our *Chandra* observations at $T_0+9$ d revealed the onset of a new emission component at X-ray energies. Although the basic model for kilonovae does not predict detectable X-ray emission, previous candidate kilonovae were all associated to an X-ray brightening. This led to the suggestion that the power source of the IR transient may be thermal re-emission of the X-ray photons rather than radioactive heat[20]. However, in these past cases[20,21,22], the X-ray luminosity was comparable or higher than the optical/IR component, while in our case the IR component is clearly dominant and 20 times brighter than the faint X-ray emission. The different luminosities and temporal behavior suggest that the X-ray emission is instead decoupled from the kilonova.

The interaction of the fast-moving ejecta with the circumstellar material may produce detectable emission[23]. An ambient density $n > 10^3$ cm$^{-3}$ would be required to explain the observed onset at $T_0+9$ d, but neither the optical nor the X-ray spectra show any evidence for absorption from this dense intervening medium. After a binary NS merger, X-rays could be produced by a rapidly rotating and highly magnetized neutron star. However, none of the current models[21,24] can reproduce persistent emission over the observed timescales of ~2 weeks. Fallback accretion[25] of the merger ejecta could account for such long-lived faint X-ray emission, however the predicted thermal spectrum should not be visible at radio frequencies. Instead, a more likely explanation, also supported by the detection of a radio counterpart, is that the observed X-rays are synchrotron afterglow radiation from the short GRB170817A. By assuming that radio and X-ray emission belong to the same synchrotron regime, we derive a spectral slope $\beta$~0.64, consistent with the index measured from the X-ray spectrum (see Methods) and with typical values of GRB afterglow spectra[15]. Therefore, our detection of X-ray emission at the same position as SSS17a (see Methods) shows that the short GRB and the optical/infrared transient are co-located, establishing a direct link between GRB170817A, its kilonova and GW170817.

In the standard GRB model[26], the broadband afterglow emission is produced by the interaction of the jet with the surrounding medium. For an observer on the jet axis, the afterglow appears as a luminous ($L_{X,iso} > 10^{44}$ erg s$^{-1}$) fading transient visible across the electromagnetic spectrum from the first few minutes after the burst. This is not consistent with our observations. If the observer is instead viewing beyond the opening angle $\theta_j$ of the jetted outflow, relativistic beaming will weaken the emission in the observer's direction by orders of magnitude. The afterglow only becomes apparent once the jet has spread and decelerated sufficiently that the beaming cone of the emission includes the observer[7,10]. Therefore, an off-axis observer sees that the onset of the afterglow is delayed by several days or weeks. In our case, the slow rise of the X-ray emission suggests that our observations took place near the peak time $t_{pk}$ of the off-axis afterglow light curve, predicted to follow $t_{pk} \propto E_{k,iso}^{1/3} n^{-1/3} (\theta_v - \theta_j)^{2.5}$, where $E_{k,iso}$ is the isotropic-equivalent blastwave energy. The off-axis angle $\Delta\theta$ is therefore constrained as $\Delta\theta = \theta_v - \theta_j \approx 13° (E_{k,iso} / 10^{50}$ erg$)^{-2/15} (n / 10^{-3}$ cm$^{-3})^{2/15}$.

In Figure 3 (panel a) we show that our dataset can be reproduced by a standard short GRB afterglow[15] with the only difference being the viewing angle: on-axis ($\theta_v \ll \theta_j$) in the commonly observed scenario, and off-axis ($\theta_v > \theta_j$) in our case. The synthetic light curves have been produced from two-dimensional jet simulations[27], but the key features of these curves are general to spreading ejecta seen off-axis (see Methods for further details; also Extended Data Figure 2). Our observations therefore independently confirm the collimated nature of GRB outflows[28].

Interestingly, all three observed electromagnetic counterparts (gamma-ray burst, kilonova and afterglow) separately point at a substantial offset of the binary orbital plane axis relative to the observer, independent of any constraint arising directly from the GW event.

The initial gamma-ray emission is unusually weak, being orders of magnitude less luminous than typical short GRBs. This suggests a significant angle between the jet and the observer. The standard top-hat profile, commonly adopted to describe GRB jets, cannot easily account for the observed properties of GRB170817A (see Methods). Instead, a structured jet profile, where the outflow energetics and Lorentz factor vary with the angle from the jet axis, can explain both the GRB and afterglow properties (Extended Data Figure 3). Alternatively, the low-luminosity gamma-ray transient may not trace the prompt GRB emission, but come from a broader collimated, mildly relativistic cocoon[29].

Another independent constraint on the off-axis geometry comes for the spectral and temporal evolution of the kilonova light curves (Figure 3, panel b). The luminous and long-lived optical emission implies that the observer intercepts a significant contribution from the wind component along the polar axis, which, for example, would be shielded by the lanthanide-rich ejecta for an edge-on observer along the equatorial plane (Figure 4). A comparison between the kilonova models[30] and our optical-infrared photometry favors an off-axis orientation, in which the wind is partially obscured by the dynamical ejecta, with an estimated inclination angle anywhere between 20° to 60° (Extended Data Figure 4), depending on the detailed configuration of the dynamical ejecta. Taking into account the uncertainties in the model, such as the morphologies of the ejecta and the possible different types of wind, this is in good agreement with the orientation inferred from afterglow modeling.

The geometry of the binary merger GW170817 (Figure 4), here primarily constrained through electromagnetic observations, could be further refined through a joint analysis with the GW signal. The discovery of GW170817 and its X-ray counterpart shows that the second generation of GW interferometers will enable us to uncover a new population of weak and likely off-axis GRBs associated with GW sources, thus providing an unprecedented opportunity to investigate the properties of these cosmic explosions and their progenitors. This paves the way for a multi-messenger modeling of the different aspects of these events, which holds the promise to play a key role in breaking the degeneracies that exist in the models when considered separately.

**Acknowledgements**

We acknowledge the advice and contribution of Neil Gehrels, who was co-Investigator of our *Chandra* and *HST* observing programs. For their support to these observations we thank Belinda Wilkes and the Chandra X-ray Center staff, Neill Reid and the STScI staff, Jamie Stevens and the CSIRO staff, Laura Ferrarese and the Gemini support staff, in particular Roberto Salinas, Morten Andersen, Hwihyun Kim, Pablo Candia, and Karleyne Silva.

ET thanks Bianca Alessandra Vekstein, Agnese Bersich, and Francesco Troja for their help and support during the composition of this manuscript. We thank Varun Bajaj (STScI) and Svea Hernandez for their assistance with data reduction. Work at LANL was done under the auspices of the National Nuclear Security Administration of the U.S. Department of Energy at Los Alamos National Laboratory under Contract No. DE- AC52-06NA25396. All LANL calculations were performed on LANL Institutional Computing resources. This research used resources provided by the Los Alamos National Laboratory Institutional Computing Program, which is supported by the U.S. Department of Energy National Nuclear Security Administration under Contract No. DE-AC52-06NA25396. MI, JK, CC, GL, and YY acknowledge the support from the NRFK grant, No. 2017R1A3A3001362, funded by the Korean government. Work by CUL and SLK was supported by the KASI (Korea Astronomy and Space Science Institute) grant 2017-1-830-03. This research has made use of the KMTNet system operated by KASI, and the data were obtained at three host sites of CTIO in Chile, SAAO in South Africa, and SSO in Australia. ET acknowledges support from grants GO718062A and HSTG014850001A. Rubén Sánchez-Ramírez acknowledges support by ASI (Italian Space Agency) through the Contract n. 2015-046-R.0 and by AHEAD the European Union Horizon 2020 Programme under the AHEAD project (grant agreement n. 654215).

**Author Information**

Reprints and permissions information is available at www.nature.com/reprints. The authors declare no competing financial interests. Correspondence and requests for materials should be addressed to eleonora.troja@nasa.gov.

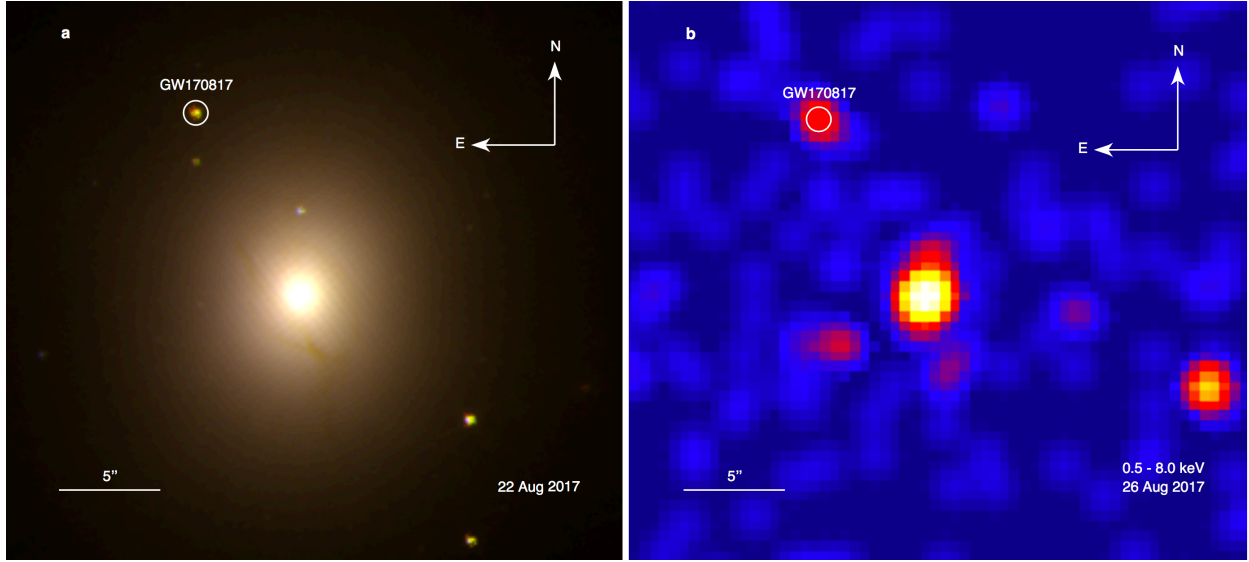

**Figure 1: Optical/Infrared and X-ray images of the counterpart of GW170817**

**a** *Hubble Space Telescope* observations show a bright and red transient in the early-type galaxy NGC 4993, at a projected physical offset of ~2 kpc from its nucleus. A similar small offset is observed in some (~25%) short GRBs[5]. Dust lanes are visible in the inner regions, suggestive of a past merger activity (see Methods). **b** *Chandra* observations revealed a faint X-ray source at the position of the optical/IR transient. X-ray emission from the galaxy nucleus is also visible.

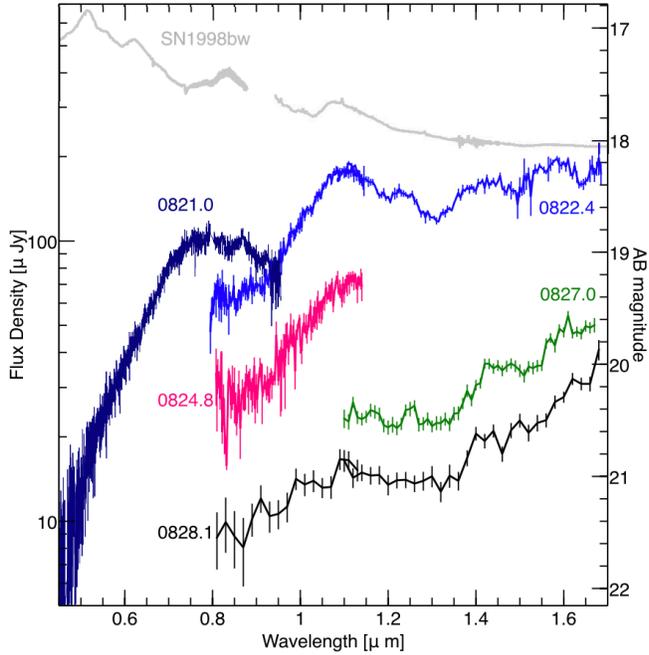

**Figure 2: Optical and infrared spectra of the kilonova associated with GW170817**

The optical spectrum, acquired on 21 Aug ($T_0$+3.5 d) with the Gemini South 8-m telescope, is dominated by a featureless continuum with a rapid turn-over above ~0.75 micron. At later times, this feature is no longer visible. Near-infrared spectra, taken with the *Hubble Space Telescope* between 22 and 28 Aug, show prominent broad ($\Delta\lambda/\lambda \approx 0.1$) features and a slow evolution toward redder colors. These spectral features are consistent with the ejection of high velocity, neutron rich material during a NS merger. A spectrum of the broad-lined Type Ic SN 1998bw (8 d post-maximum; arbitrarily rescaled) is shown for comparison. Error bars are 1 sigma.

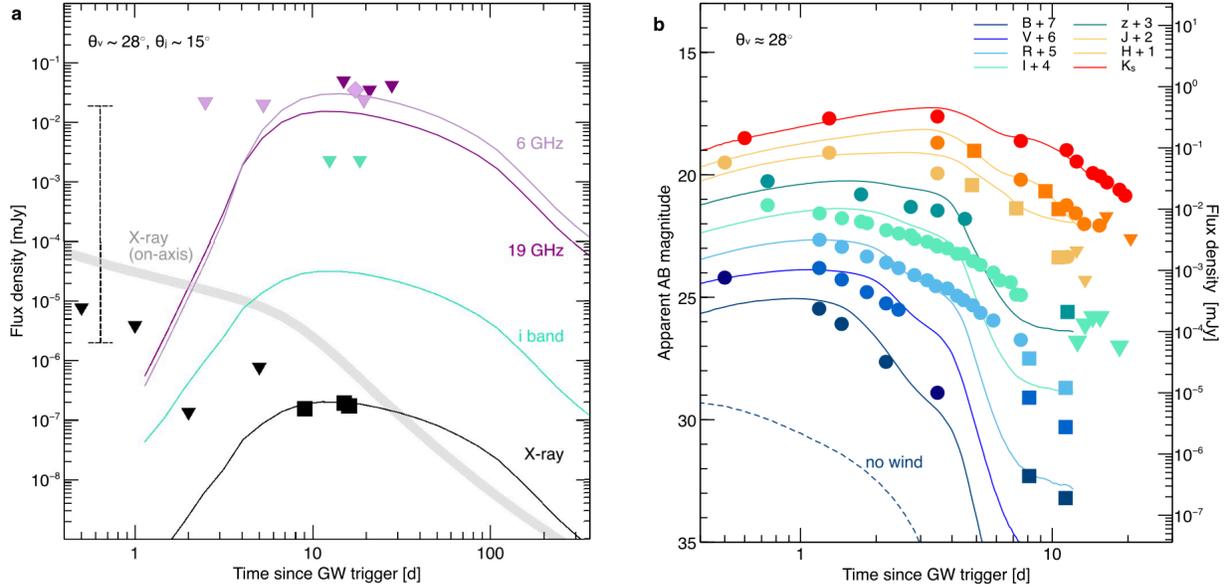

**Figure 3: Multi-wavelength light curves for the counterpart of GW170817**

**a** Temporal evolution of the X-ray and radio counterparts of GW170817 compared to the model predictions (thin solid lines) for a short GRB afterglow viewed at an angle $\theta_v \sim 28°$. The thick gray line shows the X-ray light curve of the same afterglow as seen on-axis, falling in the typical range[15] of short GRBs (vertical dashed line). Upper limits are 3 $\sigma$. **b** Temporal evolution of the optical and infrared transient SSS17a compared with the theoretical predictions (solid lines) for a kilonova seen off-axis with viewing angle $\theta_v \sim 28°$. For comparison with the ground-based photometry, *HST* measurements (squares) were converted to standard filters. Our model includes the contribution from a massive, high-speed wind along the polar axis ($M_w \sim 0.015\ M_{sun}$, $v \sim 0.08c$) and from the dynamical ejecta ($M_{ej} \sim 0.002\ M_{sun}$, $v \sim 0.2c$). The presence of a wind is required to explain the bright and long-lived optical emission, which is not expected otherwise (see dashed line).

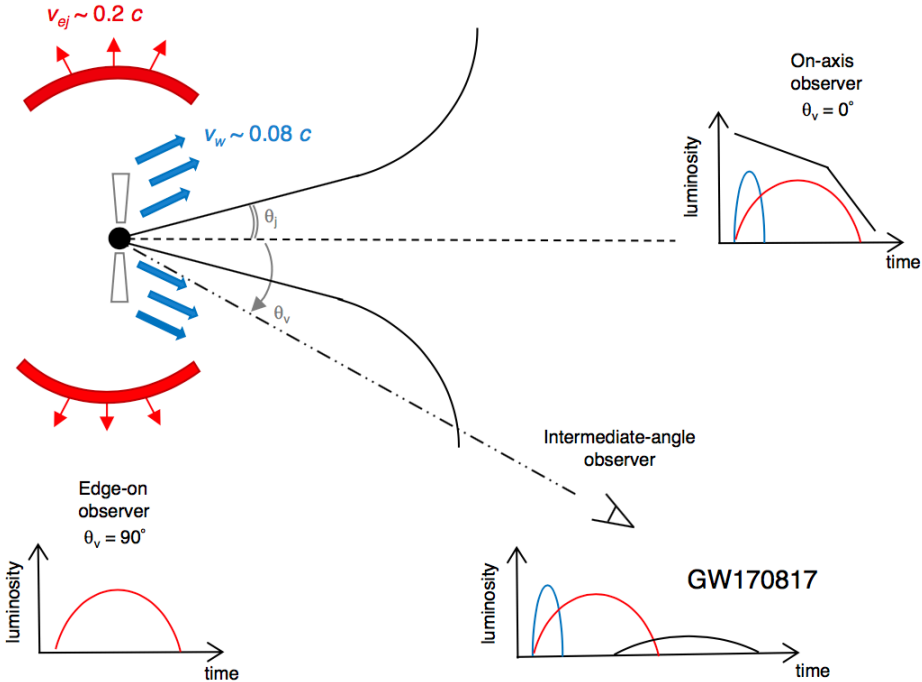

**Figure 4: Schematic diagram for the geometry of GW170817**

Following the NS merger, a small amount of fast-moving neutron-rich ejecta (red shells) emits an isotropic kilonova peaking in the infrared. A larger mass neutron-free wind along the polar axis (blue arrows) produces kilonova emission peaking at optical wavelengths. This emission, although isotropic, is not visible to edge-on observers as it is only visible within a range of angles and otherwise shielded by the high-opacity ejecta. A collimated jet (black solid cone) emits synchrotron radiation visible at radio, X-ray, and optical wavelengths. This afterglow emission outshines all other components if the jet is seen on-axis. However, to an off-axis observer, it appears as a low-luminosity component delayed by several days or weeks.

## Methods

### X-ray imaging with the Chandra X-ray Observatory

*Chandra* observed the counterpart of GW170817 at 4 different epochs. The first observation, performed at $T_0+2.2$ d, did not detect significant X-ray emission. Our observations (PI: Troja) were performed at $T_0+9$ d and $T_0+15$ d for a total exposure of 50 ks and 47 ks, respectively. Data were reduced and analyzed using standard analysis tools within CIAO v. 4.9 with calibration database CALDB v. 4.7.6. In both epochs we detect X-ray emission at the same position as the optical/IR transient (see below) at a statistically significant level (false positive probability $<10^{-7}$). The source was detected with similarly high significance in a later 47 ks observation at $T_0+16$ d.

Photon events from the afterglow were selected using a circular extraction region of radius 1 arcsec, while the background level of $2.3 \times 10^{-6}$ cts arcsec$^{-2}$ s$^{-1}$ was estimated from nearby source-free regions. In the 0.5-8.0 keV energy band, we measured 12 total counts in our first epoch and 17 total counts in the second epoch. In order to estimate the source flux, we analyzed the spectra within XSPEC. We used an absorbed power-law model with the absorbing column fixed at the Galactic value $N_H = 8.76 \times 10^{20}$ cm$^{-2}$, and minimized the Cash statistics to find our best fit parameters. The joint fit of the two spectra yielded a photon index $\Gamma = 1.3 +/- 0.4$ and unabsorbed X-ray fluxes of $(4.0 +/- 1.1) \times 10^{-15}$ erg cm$^{-2}$ s$^{-1}$ at $T_0+9$ d and $(5.0 +/- 1.0) \times 10^{-15}$ erg cm$^{-2}$ s$^{-1}$ at $T_0+15$ d in the 0.3-10 keV energy band. All the quoted errors are at the 68% confidence level (c. l.). Our results therefore suggest the presence of a slowly rising X-ray emission with $F_X \propto t^{0.5}$.

By assuming a similar background level and source spectral shape, we estimate an upper limit to the X-ray flux of $3.7 \times 10^{-15}$ erg cm$^{-2}$ s$^{-1}$ at $T_0+2.2$ d, consistent with our findings and the upper limits from *Swift* and *NuSTAR*.

**Hubble Space Telescope observations**

We obtained several epochs of imaging and near-infrared grism spectroscopy (PI: Troja) with the Hubble Space Telescope (HST). Images were taken with both the IR and the UVIS detectors of the Wide-Field Camera 3 (WFC3). Data were reduced in a standard fashion using the HST CalWF3 standard pipeline[31], and the astrodrizzle processing[32]. Fluxes were converted to magnitudes using WFC3 zero points[33,34]. Our final photometry is shown in Figure 3, panel b.

We performed relative astrometry between our WFC3/F160W image and our *Chandra* observations. We identified 5 common point-like sources (in addition to the GW counterpart SSS17a) and excluded those next to the edge of the field of view and with poor signal-to-noise. The remaining 3 sources were used to register the *Chandra* image onto the HST frame. The corrected X-ray position of SSS17a is offset from the IR position by 0.14'' +/- 0.22'' (68% c.l.). The probability of finding an unrelated X-ray source at such a small offset is $<10^{-5}$ for field objects[35] as well as for an unrelated X-ray binary within the galaxy[36]. Pre-explosion imaging[37] disfavors the presence of a globular cluster at the transient location.

Spectroscopic frames were processed with the HST CalWF3 standard pipeline. In order to estimate any possible contribution from the nearby host galaxy, we fitted a second-order polynomial (modeling the galaxy) and a Gaussian (modeling the source) as a function of the *y*-coordinate. We smoothed the resultant contamination model with a Savitzky-Golay filter to remove any high-frequency structure. We then subtracted the background and refit the remaining source flux with a Gaussian. Finally, we combined the four images (per epoch per grism) using a 3-sigma-clipped average, rejecting pixels associated with the bad-pixel masks and weighting by the inverse variance. Extended Data Figure 5 illustrates this process.

**Optical and infrared imaging with Gemini-South**

We obtained several epochs of optical and infrared imaging (PI: Troja) of the GW counterpart SSS17a, starting on 21 Aug 2017. Optical data were acquired with the Gemini Multi-Object Spectrograph (GMOS) mounted on the 8-m Gemini South telescope, and reduced using standard Gemini/IRAF tasks. We performed PSF-fitting photometry using custom Python scripts after subtracting a Sersic function fit to remove the host galaxy flux. Errors associated with the Sersic fit were measured by smoothing the fit residuals, and then propagated through the PSF fitting. The resulting *griz* photometry, shown in Figure 3 (panel b), was calibrated to Pan-STARRS[38] using a common set of field stars for all frames. Infrared images (JHKs bands) were acquired with the Flamingos-2 instrument. Data were flat-fielded and sky-subtracted using custom scripts designed for the RATIR project (http://www.ratir.org). Reduced images were aligned and stacked using SWARP. The PSF photometry was calculated, after host galaxy subtraction, and calibrated to a common set of 2MASS[39] sources, using the 2MASS zeropoints to convert to the AB system.

**Optical imaging with KMTNet**

Three Korea Microlensing Telescope Network (KMTNet) 1.6m telescopes[40] observed the counterpart of GW170817A nearly every night starting on Aug 18, 2017 at three locations, the South African Astronomical Observatory (SAAO) in South Africa, the Siding Spring Observatory (SSO) in Australia, and the Cerro-Tololo Inter-American Observatory (CTIO) in Chile. The observations were made using B, V, R, I filters. Data were reduced in a standard fashion. Reference images taken after Aug 31 were used to subtract the host galaxy contribution. Photometry was performed using SExtractor, and calibrated using the AAVSO Photometric All-Sky Survey (APASS) catalog. Our final photometry is shown in Figure 3, panel b. We also include publicly released data from Smartt, S. et al., in preparation; Pian, E. et al., in preparation; Drout, M., et al., in preparation; Chornock, R. et al. in preparation; Kasliwal, M., et al. in preparation.

**Optical Spectroscopy with Gemini**

We obtained optical spectroscopy (PI: Troja) of the GW counterpart SSS17a with GMOS beginning at 23:38 UT on 20 August 2017. A series of four spectra, each 360 s in duration, were obtained with both the R400 and B600 gratings. We employed the 1.0" slit for all observations. All data were reduced with the gemini IRAF (v1.14) package following standard procedures. The resulting spectrum of SSS17a is plotted in Figure 2. The spectrum exhibits a relatively red continuum, with a turn-over around 7500 Å. The lack of strong absorption features is consistent with the low estimated extinction along the sightline[41], $E_{B-V}$=0.105, and suggests no significant intrinsic absorption. No narrow or broad features, such as those that are typically observed in all flavors of core-collapse supernovae, are apparent.

We attempted to spectroscopically classify the source using the SuperNova IDentification (SNID) code[42], with the updated templates for stripped-envelope supernovae. No particularly good match was found, even using this expanded template set. In this case SNID often defaults to classifications of Type Ib/c (typically of the broad-lined sub-class), due to the broad (and therefore typically weaker) nature of the features. For comparison in Figure 2 we plot the spectrum of the prototypical broad-lined Type Ic supernova SN1998bw[43]. It is evident the source is not a good match. Even after removing the continuum ("flattening"), the match to mean spectral templates of broad-lined SNe Ic[44] is quite poor.

**Radio observations with ATCA**

We observed the target with the Australia Telescope Compact Array (ATCA) at three different epochs ($T_0+14.5$ d, $T_0 +20.5$ d and $T_0 + 27$ d) at the center frequencies 16.7, 21.2, 43 and 45 GHz in continuum mode (PI: Troja). The data were reduced with the data reduction package MIRIAD[45] using standard procedures. Radio images were formed at 19 and 44 GHz via the Multi Frequency Synthesis technique. No detection was found at the position of the optical/IR transient, our upper limits are shown in Figure 3, panel a. A detection of the radio afterglow at 6 GHz was reported[18] at a 5 σ level.

**Properties of the host galaxy NGC 4993**

In terms of morphology, NGC 4993 shows an extended, disturbed feature and prominent dust lanes in the inner region (Figure 1, panel a), suggestive of a minor merger in the past. From the Ks-band images we derive an absolute magnitude $M_K \sim -22$ AB mag and a stellar mass of log ($M/M_{sun}$) ~10.88, calculated by assuming a stellar mass to light ratio of order of unity[46]. Structural parameters were derived from our F110W and F160W image using GALFIT. A fit with a single Sersic component yields an index ~5.5, an ellipticity of ~0.12, and an effective radius $R_e \sim 3.4$ kpc. The lack of emission lines in our spectra suggests no significant on-going star formation at the location of the NS merger, consistent with the low UV luminosity $M_{F275W} > -7.5$ AB mag in the vicinity of the transient. Indeed, the measured Lick indices[46] with Hβ=1.23 and [MgFe]=3.16 suggest of an old (> 2 Gyr), evolved stellar population of solar or slightly sub-solar metallicity (Extended Data Figure 6). The overall properties of NGC 4993 are therefore consistent with an early-type galaxy, and within the range of galaxies harboring short GRBs[5].

In the nuclear region of NGC 4993, our radio observations show a persistent and relatively bright radio source with flux (420 +/- 30) μJy at 19 GHz. The same source is not visible at 44 GHz, indicating a steep radio spectrum. The central radio emission suggests the presence of a low-luminosity AGN contributing to the X-ray emission from the galaxy nucleus (Figure 1, panel b). AGN activity in a GRB host galaxy is rarely observed, but not unprecedented[48] in nearby short GRBs.

**Off-axis GRB modeling**

We interpret the radio and X-ray emission as synchrotron radiation from a population of shock-accelerated electrons. By assuming that radio and X-rays belong to the same synchrotron regime, we derive a spectral slope 0.64, consistent with the value measured from the X-ray spectrum $\beta = \Gamma-1 = 0.30 +/- 0.4$. This corresponds to the spectral regime between the injection frequency $\nu_m$ and the cooling frequency $\nu_c$ for a non-thermal electron population with power law index $p \sim 2.3$, close to its typical value of GRB afterglows[49]. The presence of a cooling break between radio and X-rays would imply a lower value for $p$. The apparent flattening of the X-ray light curve, and the fact that the two observations adjacent to the radio detection are upper limits, suggest that the detections were close near a temporal peak of the light curve.

We assume that the radio and X-ray detections correspond to afterglow emission from a GRB jet observed at an angle, with the observer placed at an angle $\theta_v$ outside the initial jet opening angle $\theta_j$ (Figure 4). We test two implementations of this assumption for consistency with the data, a semi-analytic simplified spreading homogeneous shell model[11] and light curves derived from a series of high-resolution two-dimensional relativistic hydrodynamics simulations[27].

Standard afterglow models[50] contain at least six free variables: $\theta_j$, $\theta_v$, isotropic equivalent jet energy $E_{iso}$, ambient medium number density $n_0$, magnetic field energy fraction $\varepsilon_B$, accelerated electron energy fraction $\varepsilon_e$. These are too many to be constrained by the observations. We therefore take 'standard' values for model parameters ($\varepsilon_B \sim 0.01$, $\varepsilon_e \sim 0.1$, $n_0 \sim 10^{-3}$, $\theta_j \sim 15°$), and choose $E_{iso}$ and $\theta_v$ to match the observations. We caution that the displayed match demonstrates only one option in a parameters space that is degenerate for the current number of observational constraints. A key feature of interest is the peak time, which is plausibly constrained by the current observations. This scales according to $t_{peak} \propto (E_{iso}/n_0)^{1/3} \Delta\theta^{2.5}$, which follows from complete scale-invariance between curves of different energy and density[51], and from a survey of off-axis curves for different using the semi-analytical model. Note that the scaling applies to the temporal *peak*, and not to the moment $t_{start}$ when the off-axis signal *starts* to become visible, where $t_{start} \propto \Delta\theta^{8/3}$ (similar to a jet break). The scaling of 2.5 is slightly shallower and reflects the trans-relativistic transition as well. $t_{peak}$ does not depend strongly on the jet opening angle, if is kept fixed. From our model comparisons to data, we infer an offset of $\Delta\theta \sim 13°$.

If a dense wind exists directly surrounding the jet, a cocoon of shocked dense material and slower jet material has been argued to exist and emerge with the jet in the form of a slower-moving outflow[52,53]. When emitted quasi-isotropically, or seen on-axis, cocoon afterglows are however expected to peak at far earlier times (~hours) than currently observed[54,55]. A more complex initial shape of the outflow than a top hat, such as structured jet[56] with a narrow core and an angle for the wings that is smaller than the observer angle, will have one additional degree of freedom. It is not possible to distinguish between the fine details of the various models: at the time of the observations, top-hat jets, structured jets and collimated cocoon-type outflows are all decelerating and spreading blast waves segueing from relativistic origins into a non-relativistic stage, and all capable of producing a synchrotron afterglow through a comparable mechanism.

**Origin of the gamma-ray emission**

For a standard top-hat GRB jet[57], the peak energy $E_p$ and the total energy release $E_{iso}$ scale as $a$ and $a^{-3}$ where $a^{-1} \sim \Gamma^2 \Delta\theta^2$ and $\Delta\theta > 1/\Gamma$. By assuming typical values of $E_{iso} \cong 2\times10^{51}$ erg, $E_p \cong 1$ MeV, and a Lorentz factor $\Gamma \cong 100$ to avoid opacity due to pair production and Thomson scattering[26,58], the expected off-axis gamma-ray emission would be much fainter than GRB170817A. This suggests that the observed gamma-rays might come from a different and probably isotropic emission component, such as precursors[59] seen in some short GRBs or a midly relativistic cocoon[54].

A different configuration is the one of a structured jet, where the energetics and Lorentz factor of the relativistic flow depend upon the viewing angle. In this case, the observed flux is dominated by the elements of the flow pointing close to the line of sight. For a universal jet, a power law dependence is assumed with $E_{\gamma,iso}(\theta_v) \propto (\theta_v/\theta_c)^{-2}$, where $\theta_c$ is the core of the jet. For a gaussian jet, the energy scales as $E_{\gamma,iso}(\theta_v) \propto \exp(-\theta_v^2/2\theta_c^2)$. Due to its significant emission at wide angles, a universal jet fails to reproduce the afterglow data (Extended Data Figure 3). A gaussian jet with standard isotropic energy $E_{\gamma,iso} \sim 2\times10^{51}$ erg can instead reproduce the observed energetics of GRB170817A ($E_{\gamma,iso} \sim 5 \times 10^{46}$ erg) when $\theta_v \sim 4\theta_c$. The same jet can also describe the broadband afterglow data, thus representing a consistent model for the prompt and afterglow emissions.

**Kilonova modelling**

Our kilonova (or macronova) calculations are based on the approach developed by [30]. We use the multigroup, multidimensional radiative Monte Carlo code *SuperNu*[62-64] (https://bitbucket.org/drrossum/supernu/wiki/Home) with the set of opacities produced by the Los Alamos suite of atomic physics codes[65-67]. For this paper, we build upon the range of two-dimensional simulations[30] using the class "A" ejecta morphologies and varying the ejecta mass, velocity, composition and orientation as well as the model for the energy deposition in post-nucleosynthetic radioactive decays. Our nuclear energy deposition is based on the finite-range droplet model (FRDM) of nuclear masses.

Kilonova light-curves can be roughly separated into two components: an early peak dominated by the wind ejecta (where by "wind" we indicate the entire variety of secondary post-merger outflows, with many elements in the atomic mass range between the iron peak up through the second r-process peak) and a late IR peak that is powered by the lanthanide-rich (main r-process elements) dynamical ejecta. The luminous optical and UV emission require a large wind mass ($M_w$>0.015-0.03 $M_{sun}$) and a composition with moderate neutron richness ("wind 2" with $Y_e$=0.27 from [30]). A large fraction of these ejecta is 1st peak r-process elements. The late-time IR data probe the properties of the dynamical ejecta ($Y_e$<0.2), arguing for a mass of $M_{ej} \sim 0.001 - 0.01\ M_{sun}$. This ejecta is primarily composed of the main r-process elements lying between the 2nd and 3rd r-process peaks (inclusive). Within the errors of our modelling, the low inferred ejecta mass combined with the high rate of neutron star mergers inferred from this GW detection is in agreement with the neutron star mergers being the main site of the r-process production[68]. However, our models seem to overproduce the 1st peak r-process relative to the 2nd and 3rd peaks. This could be due to the model simplifications in the treatment of ejecta composition, or this particular event is not standard for neutron star mergers.

Another, more plausible source of error, comes from the uncertainties in nuclear physics, such as the nuclear mass model used in the r-process nucleosynthesis calculation. Our baseline nuclear mass model (FRDM[69]) tends to underestimate the nuclear heating rates, compared to other models, e.g. DZ31 model[70]. Specifically, in the latter model the abundances of trans-lead elements can dramatically alter the heating at late times[68,71]. Combined differences in the heating rate and thermalization translate to nearly a factor of 10 in the nuclear energy deposition at late times[71] ($t$>2 days). We have therefore adjusted the heating rate in the dynamical ejecta to compensate for this effect. If this nuclear heating rate is too high, then we are underestimating the mass of the dynamical ejecta.

The opacity of the lanthanide-rich tidal ejecta is dominated by a forest of lines up to the near infrared, causing most of the energy to escape beyond 1 micron and one indicator of an ejecta dictated by lanthanide opacities is a spectrum peak above 1 micron that remains relatively flat in the IR. However, standard parameters for the ejecta predict a peak between 5-10 d. To fit the early peak (~3 d) requires either a lower mass, or higher velocities. Our best fit model has a tidal/dynamic ejecta mass of $M_{ej} \sim 0.002\ M_{sun}$. and median velocity (~$v_{peak}/2$) of 0.2c.

Extended Data Figure 4 shows our synthetic light curves for different viewing angles. In the on-axis orientation, the observer can see both types of outflows, while in the edge-on orientation the wind outflow is completely obscured. The system orientation most strongly affects the behaviour in the blue optical bands, while the infrared bands are largely unaffected. The observed slow decline in the optical bands for this event is best fit by moderate-latitude viewing angles (~20-60 degrees).

**Additional References**

**Data availability:** All relevant data are available from the corresponding author upon reasonable request. Data presented in Figure 3 (panel b) are included with the manuscript.

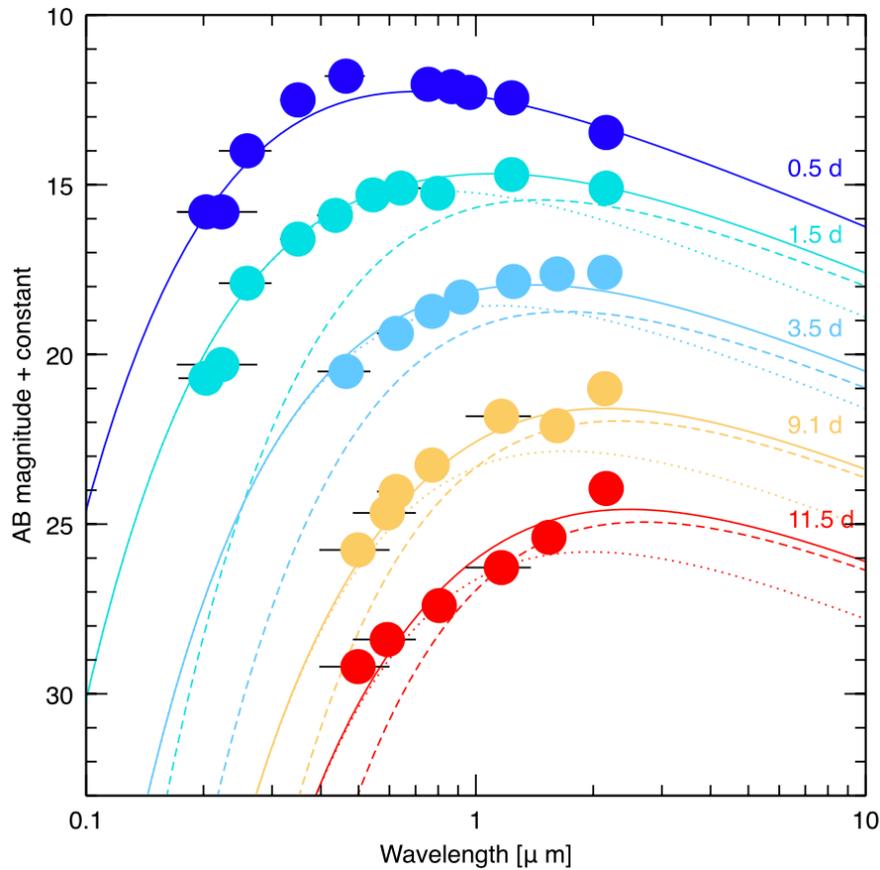

**Extended Data Figure 1 - Spectral energy distributions of the optical/infrared counterpart**

We can empirically describe the spectral energy distribution and its temporal evolution as the superposition of two blackbody components in linear expansion. A single component provides a good fit at early times ($T_0$+0.5 d), but at later times we find that two components (shown by the dashed and dotted lines) with different temperatures and expansion velocities represent a better description of the dataset. The large effective radii ($R > 4 \times 10^{14}$ cm at $T_0$+0.5 d) inferred from the blackbody fits imply an average velocity $v > 0.2\,c$. Magnitudes are corrected for Galactic extinction along the line of sight[41]. Data have been shifted for plotting purposes.

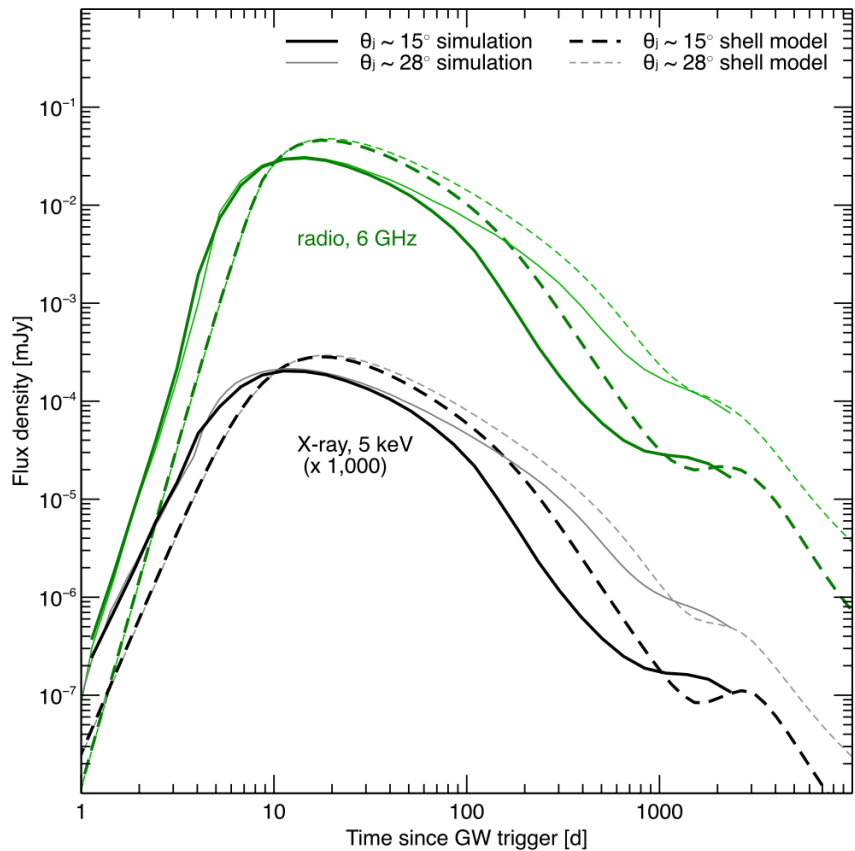

**Extended Data Figure 2 - Models of off-axis afterglows at X-ray and radio energies**

Direct comparison between off-axis light curves for two different jet opening angles (15º and 28º). As long as the difference between the viewing angle and the jet angle is maintained, a continuous range of jet angles can be found consistent with the observations in X-rays and at radio wavelengths observations mostly covering the peak. Dashed lines show light curves computed using the semi-analytic spreading top-hat jet model[11] for identical input parameters. Note that the simulated angular fluid profile quickly becomes complex as the jet evolves, and the similarity in light curves to those derived from the top-hat shell illustrate that the global features do not depend strongly on this angular profile. The simulated light curves include synchrotron self-absorption, which was not found to play an important role for the current parameters.

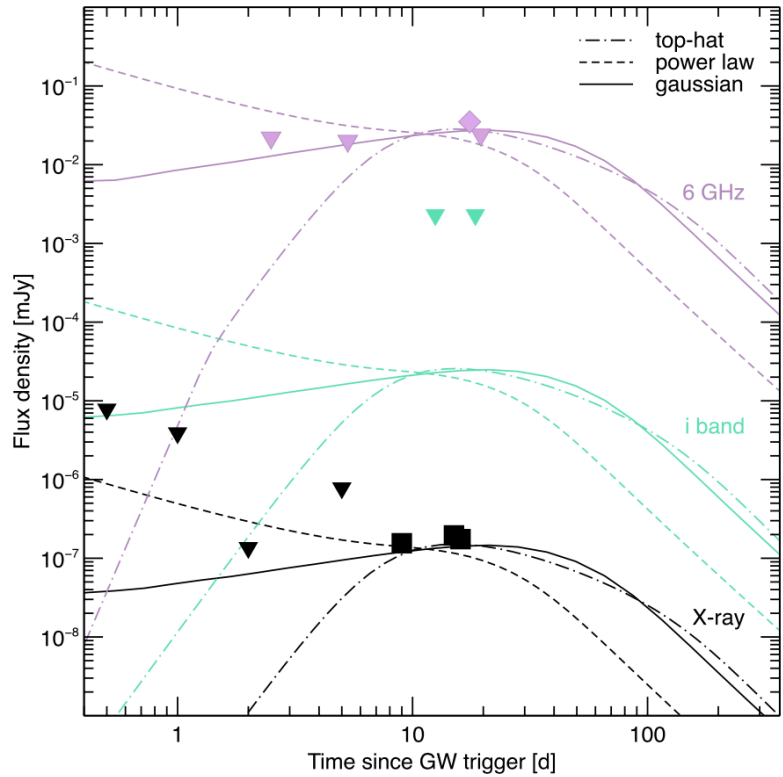

**Extended Data Figure 3 – Afterglow modeling for different jet profiles viewed at an angle**

We consider three well-known jet profiles: top-hat (dot-dashed line), gaussian (solid line), and power law (dashed line). A power law structured jet is not consistent with the lack of afterglow detection at early times. A top-hat jet and a gaussian structured jet can describe the afterglow behavior, and imply a significant off-axis angle. The gaussian jet has the additional advantage of consistently explaining both the prompt gamma-rays and the afterglow emission.

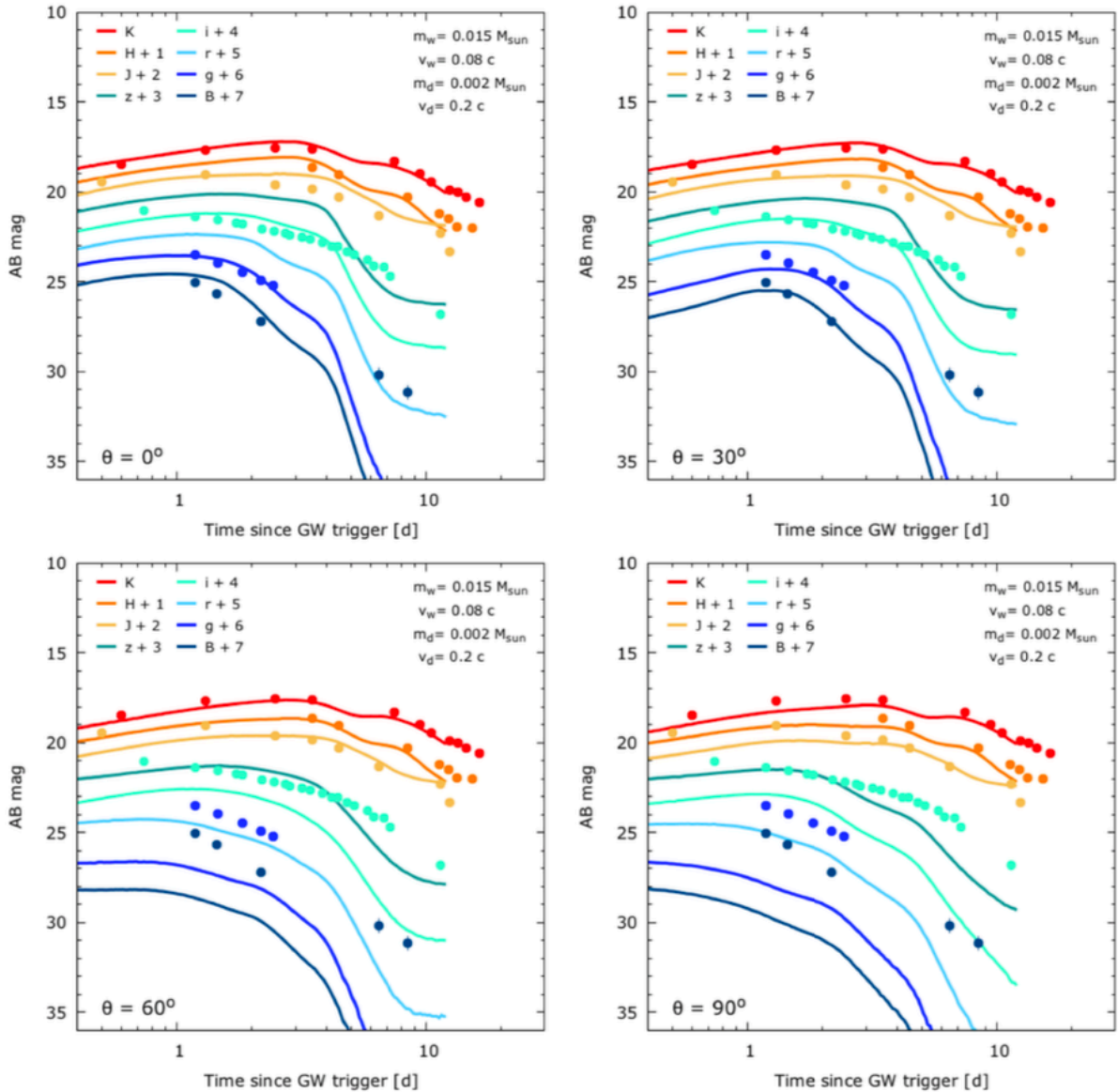

**Extended Data Figure 4 - Kilonova light curves as a function of the viewing angle**

Comparison of the observational data with the synthetic light curves from the two-component axisymmetric radiative transfer model at different viewing angles: 0 deg (a; on-axis view); 30 deg (b), 60 deg (c) and 90 deg (d; edge-on equatorial view).

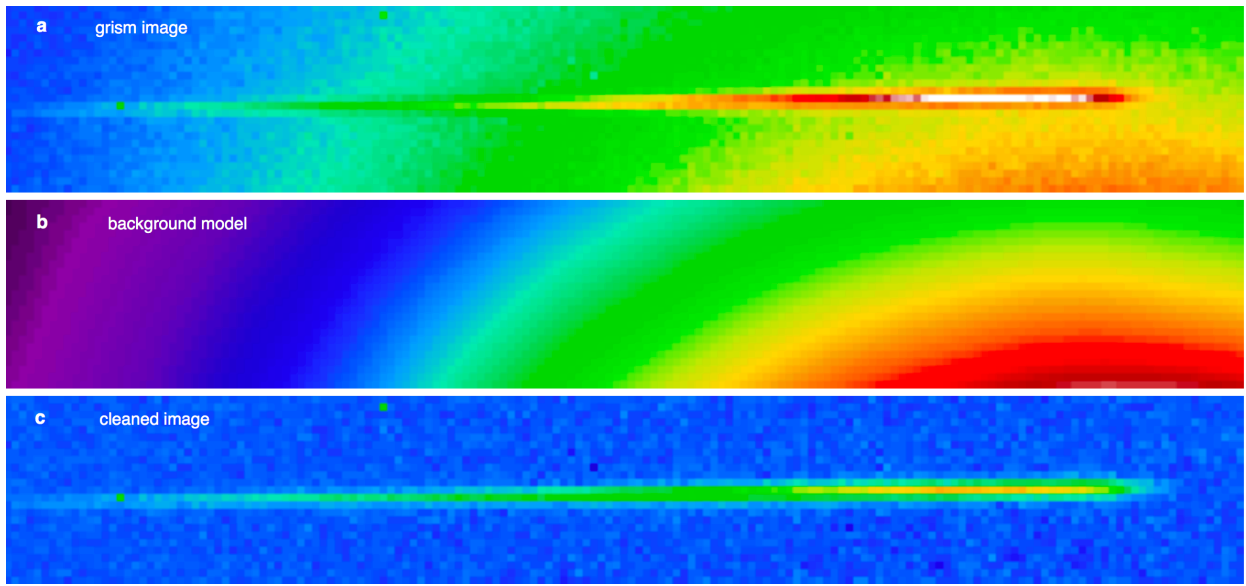

**Extended Data Figure 5 - Illustrative example of the contamination modeling**.

**a** Two-dimensional dispersed image at the position of SSS17a. **b** Our model describing the emission from NGC4993, smoothed with a Savitzky-Golay filter in order to remove any high-frequency structure. **c** Difference between the data and the model.

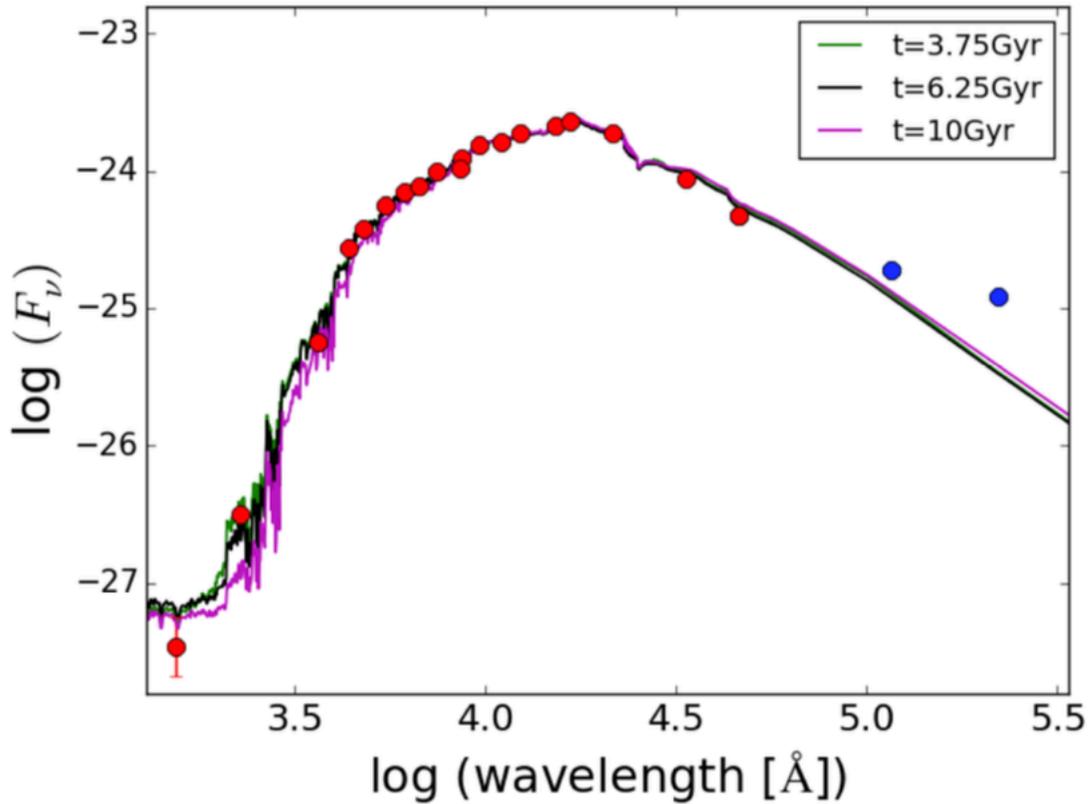

**Extended Data Figure 6 - Broadband spectral energy distribution of NGC 4993**

The model assumes a delayed star formation rate, standard spectral templates[72] and initial mass function[73]. Models for three different stellar ages are shown. Data above 5 μm are not used in the fit as they may be affected by emission from dust. The SED-fitting result prefers a mean stellar age of 3-7 Gyr and disfavors ages less than 2 Gyr. The mean stellar mass is found to be in the range of (5-10) x $10^{10}$ $M_{sun}$ with a solar metallicity.